\documentclass[aps,prb,showpacs,foatfix,nobibnotes,footinbib,
               superscriptaddress,preprint]{revtex4}

\usepackage{amsmath,amssymb,amsfonts}
\usepackage{graphicx}
\usepackage[active]{srcltx} % DVI Inverse Search

\newcommand{\nn}{\nonumber}
\newcommand{\ev}[1]{\langle {#1} \rangle}

     \newcommand{\be}{\beta}
     \newcommand{\de}{\delta}
   \newcommand{\ve}{\varepsilon}

\newcommand{\la}{\lambda}    
   
  \newcommand{\te}{\theta}

\newcommand{\De}{\Delta}

 \newcommand{\cP}{\mathcal{P}}

\begin{document}

\title{Competing effects of interactions in the adsorbed and activated states on surface diffusion at low temperatures}

\author{Igor Medved'}
\email{imedved@ukf.sk}
\affiliation{Department of Physics, Constantine the Philosopher University, 94974 Nitra, Slovakia}
\affiliation{Department of Materials Engineering and Chemistry, Czech Technical University, 16629 Prague, Czech Republic}
\author{Anton Trn\'{\i}k}
\affiliation{Department of Physics, Constantine the Philosopher University, 94974 Nitra, Slovakia}
\affiliation{Department of Materials Engineering and Chemistry, Czech Technical University, 16629 Prague, Czech Republic}
\author{Robert \v Cern\'y}
\affiliation{Department of Materials Engineering and Chemistry, Czech Technical University, 16629 Prague, Czech Republic}

% ----------------------------------------------------------------
\begin{abstract}
The competing influence of two types of interactions on surface diffusion is investigated: one of them, $\ve$, acts between adsorbed particles, while the other one, $\eta$, between activated and adsorbed particles. To this end, a specific lattice-gas model on a triangular, square, and hexagonal lattice is considered with an attractive $\ve$ and an attractive or repulsive $\eta$, both restricted to nearest neighbors. For all three lattices the influence is qualitatively the same. Namely, when $\eta$ is neglected, then $\ve$ is shown to decelerate diffusion with a rate exponential in $\ve$ and inverse temperature. Moreover, when $\eta$ is present and $\ve$ is fixed, then a sufficiently attractive $\eta$ (relative to $\ve$) accelerates diffusion exponentially fast in $\eta$ and inverse temperature. However, quite surprisingly, a repulsive or slightly attractive $\eta$ has practically no effect on diffusion. Finally, when $\eta$ is set proportional to $\ve$ via a parameter $a$, surface diffusion is exponentially accelerated (decelerated) for $a$ above (below) a threshold value equal to $4$, $5/4$, and $7/8$ for the three lattices. Thus, an $\eta$ of strength comparable to $\ve$ is enough to boost diffusion on the square and hexagonal lattices, while an $\eta$ of a rather large strength is needed for the boost on the triangular lattice.
\end{abstract}


\maketitle

% ----------------------------------------------------------------
% ****************************************************************
% ----------------------------------------------------------------

% ######################################################## SECTION

\section{Introduction} \label{sec: INT}

The importance of surface diffusion as a mechanism of mass transport has been acknowledged in various fields, such as chromatography, \cite{MiGu10} sintering, \cite{shi05,Cha08} microelectronics, \cite{Tu03,Kax96} civil engineering, \cite{Cer11,MC11} heterogeneous catalysis, \cite{Kat04} or neuroscience. \cite{New08,Gro09} A widely used approach to simulate surface diffusion at a microscopic level is to employ lattice-gas models. \cite{Ala02,Ev06} In these models the diffusion is given by a potential relief of the surface. While most of the time the particles stay at adsorption sites (minima of the relief), occasionally they perform random jumps to adjacent vacant sites. If the jumps are fast enough, the microstates of the system can be represented by occupation numbers (one number for each site), as in a lattice gas.

Interactions between particles may strongly influence surface diffusion. Indeed, attractive interactions between particles adsorbed on the surface restrain their migration and, thus, slow down diffusion. By the same token, repulsion between adsorbed particles boosts diffusion. This basic picture is obscured by the fact that particles in the activated state---those performing jumps between two sites and located at saddle points of the potential relief---also interact with the adsorbed particles within their vicinity, which may have a significant impact on diffusion as well. However, the effect of this interaction is opposite: if it is attractive, it tends to lower the effective barriers of the jumps, which leads to higher jump probabilities and a boost in diffusion; if it is repulsive, it should slow diffusion down. Using Monte Carlo simulations at temperatures above the critical point, the effect of each of these two types of interactions on diffusion was already studied separately, corroborating the above-described anticipated behavior. \cite{Ta01,Ta00} A combined effect of both types of interaction has not been studied, though. Hence, it still remains unclear which effect is dominant and under what conditions. This point is investigated in detail in the present paper.

To this end, we employ a simple lattice-gas model with an attractive nearest-neighbor interaction, $\ve$, between adsorbed particles, and a different (attractive or repulsive) interaction, $\eta$, between an activated particle and its closest adsorbed particles. A triangular, square, and hexagonal lattice will be considered, all of which will turn out to exhibit the same type of behavior in surface diffusion coefficients.

Our analysis is based on two basic assumptions. First, we will assume that the surface coverage varies only very slowly with time and space, i.e., that the local equilibrium approximation is applicable. Then the chemical and jump diffusion coefficients, $D_c$ and $D_J$, can be expressed via purely thermodynamic quantities. As various comparisons with kinetic simulations have revealed, the results obtained within this approximation are reliable even farther away from equilibrium, \cite{Ta03,Ta07a,Ta07b,Ta09} indicating that the approximation is actually quite robust. Second, we assume that the temperature is low (sufficiently below the critical point) so that ordered phases of adsorbed particles occur in the system. Then explicit formulas for coverage dependences of $D_c$ and $D_J$ are available, \cite{MT12a} and the needed thermodynamic quantities can be evaluated (to any desired precision, in principle) via cluster expansions. In particular, one can evaluate various multi-site correlation functions that inevitably arise due to the presence of an activated-state interaction. \cite{MT13} Previously, the correlations were obtained only by approximative techniques, like the cluster variation method \cite{Da96,Da98} or approximations via two- or three-site correlations. \cite{Ta00,Ta01} Rather crudely, the activated-state interactions were often neglected to work only with the correlations between nearest-neighbor sites. \cite{Zh85,Zh91,Zh95,Ta03,Ta07a,Ta12}

The paper structure is as follows. In Section~\ref{sec: MODEL} we introduce the studied lattice-gas model and give expressions for the coefficients $D_c$ and $D_J$ in terms of the surface coverage, isothermal susceptibility, and a correlation factor. In Section~\ref{sec: RES} we then investigate the effects of the two interactions $\ve$ and $\eta$ on these coefficients, employing general formulas for the coverage dependence of $D_c$ and $D_J$. Upon evaluating the quantities involved in these formulas for our model, we obtain explicit dependences of $D_c$ and $D_J$ on the interactions, allowing us to analyze in detail their influence on surface diffusion. A summary of our results and concluding remarks are given in a final section.

% ####################################################### SECTION
\section{Model of surface diffusion}
\label{sec: MODEL}

We shall consider an ideal solid surface whose potential relief minima form a triangular, square, or hexagonal lattice. Each lattice site, $x$, is either vacant (an occupation number $\nu_x = 0$) or occupied by an adparticle ($\nu_x = 1$). An interaction between adparticles is assumed to occur only between nearest neighbors (\emph{nn}s), the corresponding interaction energy being attractive, $\ve < 0$. The model Hamiltonian is given as
\begin{equation} \label{eq: H}
  H = \ve N_2 - \mu N_1,
\end{equation}
where $N_1$ ($N_2$) is the number of occupied sites (occupied \emph{nn} pairs) in a given microstate and $\mu$ is the chemical potential.

The model is equivalent to the Ising model of ferromagnetism and was already used to simulate surface diffusion at high temperatures $T$. \cite{UG95,Ta00,Ta01,Ta03}  In particular, it was employed to study the effects of an attractive interaction $\ve$ on surface diffusion on a triangular lattice (for $T \geq 0.94 |\ve|/k_B$) and on a square lattice (for $T \geq 0.568 |\ve|/k_B$ and $\ve$ attractive). \cite{Ta03,Ta00} Note that these temperatures are above the critical values $0.9102 |\ve|/k_B$ and $0.5673 |\ve|/k_B$ for the two lattices (for a hexagonal lattice the critical point is $0.3797 |\ve|/k_B$). \cite{Pat11}

At low temperatures model~\eqref{eq: H} is known to have only two phases: a fully vacant phase at $\mu < \mu_t$ and a fully occupied phase at $\mu > \mu_t$; at $\mu = \mu_t$ both phases coexist. \cite{Gal99,HM02} Thus, the models exhibits a first-order transition at $\mu_t$ between the fully vacant and fully occupied phase. The point $\mu_t = q\ve/2$, where $q$ is the lattice coordination number (equal to $6$, $4$, and $3$ for the three lattices).

\begin{figure}
  \centering
  \includegraphics{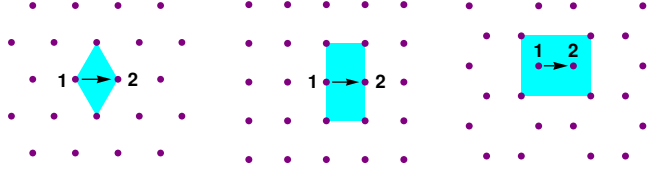}
  \caption{The three lattices with a jump between a pair of \emph{nn} sites $1$ and $2$. The segments $S$ are indicated by shaded areas.}
  \label{fig: segm}
\end{figure}

Migration due to surface diffusion is represented by jumps of adparticles to nearby vacant sites. For simplicity, we shall assume that these jumps are uncorrelated and occur only between \emph{nn} sites. The influence of adparticles on the activation energy of a jump will be taken into account via an (attractive or repulsive) interaction, $\eta$, that is different from $\ve$ and acts between an activated particle and its closest adparticles. \cite{Ta00,Ta01} Thus, this interaction is associated with segments, $S$, of more than two adsorption sites (see Fig.~\ref{fig: segm}): a pair of \emph{nn} sites between which a particle jump is performed, plus the closest sites to the pair's center (over which a saddle point of the potential relief is located).

Relevant many-particle transport parameters for surface diffusion are the chemical and jump diffusion coefficients, $D_c$ and $D_J$, respectively. The former coefficient describes surface mass-transport processes of the system and is defined via the Fick's first law; the latter describes the asymptotic behavior of the mean square displacement of the system center of mass. They are related by the Kubo-Green equation, $D_c = (\be \te / \chi) D_J$, \cite{Zh95,Re81a} where $\te$ is the surface coverage, $\chi$ is the isothermal susceptibility, and $\be = 1/k_B T$ is the inverse temperature.

As long as the coverage varies only very slowly with time and space (i.e., the local equilibrium limit is applicable), purely thermodynamic quantities are sufficient to approximate the two diffusion coefficients, \cite{Tar80,Re81a,Re81,Zh91,Ta01,Ta03}
\begin{equation} \label{eq: D gen}
  D_c \approx D_0 e^{\mu\be} \frac P{\chi/\be},
  \qquad
  D_J \approx D_0 e^{\mu\be} \frac P\te,
\end{equation}
where $D_0$ is the diffusion coefficient of non-interacting particles.

\begin{figure}
  \centering
  \includegraphics{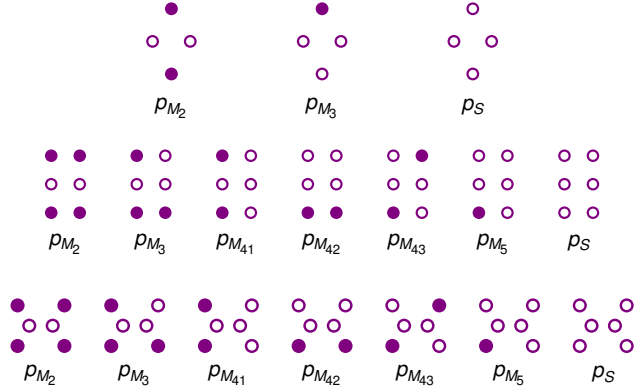}
  \caption{The statistical averages $p_M$ of finding holes at the sites of $M$ (indicated by circles) and particles at the other sites of $S$ (indicated by disks) for the three lattices. A given $p_M$ may correspond to several sets $M$ of which only one is shown (for example, there are two sets $M_3$ on a triangular lattice associated with $p_{M_3}$).}
  \label{fig: prob}
\end{figure}

The correlation factor $P$ is associated with the interaction of activated particles and may be written as \cite{Ta00,Ta01}
\begin{equation} \label{eq: P}
  P = \sum_M e^{- (n_S - n_M) \eta \be} p_M.
\end{equation}
The summation is over all subsets $M$ of a segment $S$ that contain the pair of \emph{nn} sites between which a jump is performed, including $S$ itself (see Fig.~\ref{fig: prob}), and $n_M \geq 2$ is the number of sites in $M$. Moreover, $p_M$ is the statistical average that the sites of $M$ are vacant and the remaining sites in $S$ are occupied. For example, for a triangular lattice and $M = M_3$ we have $n_S = 4$, $n_{M_3} = 3$, and $p_{M_3} = \ev{(1-\nu_1) (1-\nu_2) (1-\nu_3) \nu_4}$, where $1, \dots, 4$ are the sites of $S$ and $1$ and $2$ are the \emph{nn} sites associated with a particle jump. Taking into account that there may be several sets $M$ yielding the same value of the average $p_M$, one explicitly has \cite{Ta00,Ta01}
\begin{subequations} \label{eq: P expl}
\begin{align}
  P &= e^{-2\eta\be} p_{M_2} + 2 e^{-\eta\be} p_{M_3} + p_S
\\
  \intertext{for the triangular lattice and}
  P &= e^{-4\eta\be} p_{M_2} + 4 e^{-3\eta\be} p_{M_3}
  \nn
\\
  &\quad
  + 2 e^{-2\eta\be} p_4 + 4 e^{-\eta\be} p_{M_5} + p_S
\end{align}
\end{subequations}
for the square and hexagonal lattices, where the shorthand $p_4 = p_{M_{41}} + p_{M_{42}} + p_{M_{43}}$.

% ######################################################## SECTION
\section{Results and discussion} \label{sec: RES}

According to Eq.~\eqref{eq: D gen}, in the local equilibrium limit the diffusion coefficients $D_c$ and $D_J$ become thermodynamic quantities. A rather powerful statistical mechanical machinery \cite{BoKo90,BoKo95,Zah84,BI89} was developed to study such quantities at low temperatures. We shall apply it here to investigate the coverage dependences of $D_c$ and $D_J$, focusing on their variations with the interactions $\ve$ and $\eta$. Before doing so, let us mention two peculiar features of systems at low temperatures that underlie the forthcoming results.

The first one is the structure of phases. At low temperatures a typical microstate in a given phase looks as a `sea' of a ground state in which isolated `islands' of non-ground-state configurations are scattered. \cite{Zah84,BI89,MHTV13} For the fully vacant (occupied) phase in model~\eqref{eq: H} this corresponds to a sea of holes (particles) with islands of particles (holes). The islands are small, with diameters of order not exceeding $\ln N$, yet their bulk density is non-zero, of order $\exp(- c\be) \ll 1$. In fact, the smallest islands (of one or few sites) yield the dominant contributions to the system's free energy. As a result, in a given phase the coverage has a practically constant value corresponding to the ground state associated with the phase (a value $0$ or $1$ for model~\eqref{eq: H}), the deviations being just of order $\exp(- c\be)$. Hence, at low temperatures a significant change in the coverage can possibly occur only at or very near a transition between phases.

The second feature is the smoothing of a discontinuity in a thermodynamic quantity (such as the coverage) at a first-order phase transition in a finite system. At low temperatures the profile of the smoothing is identical (given by the function $\tanh$) for a large group of lattice-gas models. \cite{BoKo90,BoKo95} This allows one to described the dependence of thermodynamic quantities on the coverage in a unified way.

% ===============================================================
\subsection{Formulas for diffusion coefficients}

These facts can be used to obtain general low-temperature formulas for the coverage dependences of $D_c$ and $D_J$ associated with a first-order phase transition between two phases. \cite{MT12a} In the case of model~\eqref{eq: H} they read
\begin{equation} \label{eq: D theor}
\begin{aligned}
  D_c &\approx \frac{ D_0 e^{\mu_t\be} }{ \De\te \, N }
  \Bigl( \frac{P_\circ}{\te - \te_\circ}
  + \frac{P_\bullet}{\te_\bullet - \te} \Bigr),
\\
  D_J &\approx \frac{ D_0 e^{\mu_t\be} }{ \De\te }
  \Bigl(  P_\circ \, \frac{\te_\bullet - \te}\te
        + P_\bullet \, \frac{\te - \te_\circ}\te \Bigr),
\end{aligned}
\end{equation}
where $N$ is the total number of adsorption sites in the system, and the constants $\te_\circ, \te_\bullet$ and $P_\circ, P_\bullet$ are the coverages and correlation factors, respectively, evaluated at the transition point $\mu_t$ in the fully vacant (symbol `$\circ$') and fully occupied (symbol `$\bullet$') phase. The shorthand $\De\te = \te_\bullet - \te_\circ$ is the coverage discontinuity at the transition.

\begin{figure}
  \centering
  \includegraphics{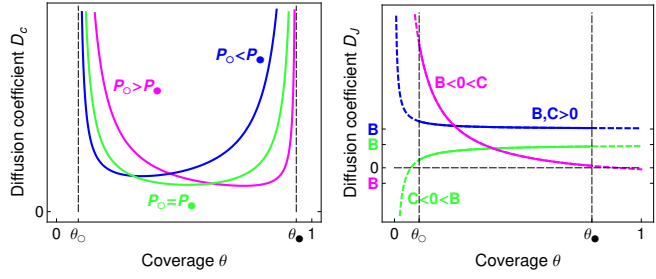}
  \caption{The coverage dependence of the diffusion coefficients as given by Eq.~\eqref{eq: D theor} for various possibilities of parameters values. For the jump coefficient $D_J \approx B + C/\te$ the parameters $B = P_\bullet - P_\circ$ and $C = P_\circ \te_\bullet - P_\bullet \te_\circ$ cannot be simultaneously negative.}
  \label{fig: D gen}
\end{figure}

Formulas~\eqref{eq: D theor} work for a system with periodic boundary conditions and for coverages $\te_\circ < \te < \te_\bullet$. This is almost the whole interval $0 < \te < 1$, as $\te_\circ \approx 0$ and $\te_\bullet \approx 1$ at low temperatures (see Eq.~\eqref{eq: te12} below). The formulas show that the coefficient $D_c$ behaves as a sum of two hyperbolas diverging at $\te_\circ$ and $\te_\bullet$, respectively, and the coefficient $D_J$ as a hyperbola $B + C/\te$ diverging at $0$ (see Fig~\ref{fig: D gen}). No divergences in $D_c$ or $D_J$ actually occur, however, because for $0 \leq \te \leq \te_\circ + \de$ and $\te_\bullet + \de \leq \te \leq 1$ with $\de \sim N^{-3/4} \ll 1$ one of the two phases prevails in the system, and formulas different from Eq.~\eqref{eq: D theor} become valid. \cite{MT12a} Nevertheless, in the following we shall use only Eq.~\eqref{eq: D theor} since it is applicable to practically all coverages.

The interactions $\ve$ and $\eta$ affect the diffusion coefficients via the parameters $P_\circ, P_\bullet$, $\te_\circ, \te_\bullet$, $\De\te$, and $\mu_t$  that we shall now evaluate. Note that $\eta$ appears only in the correlation factors $P_\circ, P_\bullet$ (via the exponential $\exp(-\eta\be)$), whereas the interaction $\ve$ affects $P_\circ, P_\bullet$ (via the averages $p_M^\circ, p_M^\bullet$) as well as the coverages $\te_\circ, \te_\bullet$ and location of the transition point $\mu_t$.

% ===============================================================
\subsection{Single-phase coverages and correlation factors}

At low temperatures, the leading contributions to the free energy in a given phase come from single-site thermal perturbations of the ground state. In the vacant phase such a perturbation corresponds to microstate with an energy excess $\De H = -\mu > 0$ over the vacant ground state, while in the occupied phase the excess is $\De H = \mu - q\ve > 0$. These excesses correspond to the Boltzmann factors $\exp(\mu\be)$ and $\exp[(q\ve-\mu)\be]$, respectively. Hence, \cite{MT12,MT13}
\begin{equation} \label{eq: te12}
\begin{aligned}
 \te_\circ &\approx e^{\mu_t\be} = e^{q\ve\be/2},
 \\
 \te_\bullet &\approx 1 - e^{(q\ve - \mu_t)\be}
 = 1 - e^{q\ve\be/2}.
\end{aligned}
\end{equation}
Note that this yields $\De\te \approx 1 - 2 \exp(q\ve\be/2)$.

In an analogous way, we may obtain the single-phase averages $p_M^\circ$ and $p_M^\bullet$ at the transition. Consider, for example, the average $p_{M_{42}}^\circ$ on a square lattice. It corresponds to addition of two \emph{nn} adparticles to the fully vacant ground state (see Fig.~\ref{fig: prob}). Similarly, $p_{M_{42}}^\bullet$ corresponds to removal of four adparticles in an elementary square from the fully occupied ground state. The energy excesses of these microstates are $\De H = \ve - 2\mu$ and $\De H = 4\mu - 12\ve$, respectively. Thus, within the leading-order approximation, $p_{M_{42}}^\circ \approx \exp[(2\mu_t-\ve)\be] = \exp(3\ve\be)$ and $p_{M_{42}}^\bullet \approx \exp[(12\ve-4\mu_t)\be] = \exp(4\ve\be)$. In this way the leading terms in all averages $p_M^\circ$ and $p_M^\bullet$ can be deduced (see Table~\ref{tab: p_M}). Although this evaluation is heuristic, it can be shown to be actually correct. \cite{MT13} It must be taken into account, however, that there may be two or three different types of perturbations of a ground state corresponding to the leading term in $p_M^\circ$ or $p_M^\bullet$, yielding a non-unit multiplicative prefactor in such a case.
\begin{table*}
\centering
\begin{ruledtabular}
\begin{tabular}{lllllllll}
  & \multicolumn{2}{c}{Triangular lattice} &
  & \multicolumn{2}{c}{Square lattice} &
  & \multicolumn{2}{c}{Hexagonal lattice}
\\ \cline{2-3} \cline{5-6} \cline{8-9}
  $M$ & $p_M^\circ$ & $p_M^\bullet$ &
   & $p_M^\circ$ & $p_M^\bullet$ &
   & $p_M^\circ$ & $p_M^\bullet$
\\ \hline
  $M_2$ & $e^{6\ve\be}$ & $e^{5\ve\be}$ &
  & $e^{6\ve\be}$ & $e^{3\ve\be}$ &
  & $4 e^{6\ve\be}$ & $e^{2\ve\be}$
\\
  $M_3$ & $e^{3\ve\be}$ & $e^{6\ve\be}$ &
  & $e^{5\ve\be}$ & $e^{4\ve\be}$ &
  & $2 e^{9\ve\be/2}$ & $e^{5\ve\be/2}$
\\
  $M_{41}$ & -- & -- &
  & $e^{4\ve\be}$ & $e^{5\ve\be}$ &
  & $e^{3\ve\be}$ & $e^{3\ve\be}$
\\
  $M_{42}$ & -- & -- &
  & $e^{3\ve\be}$ & $e^{4\ve\be}$ &
  & $2 e^{3\ve\be}$ & $2 e^{3\ve\be}$
\\
  $M_{43}$ & -- & -- &
  & $e^{4\ve\be}$ & $e^{5\ve\be}$ &
  & $e^{3\ve\be}$ & $e^{3\ve\be}$
\\
  $M_5$ & -- & -- &
  & $e^{2\ve\be}$ & $e^{5\ve\be}$ &
  & $e^{3\ve\be/2}$ & $2 e^{7\ve\be/2}$
\\
  $S$ & $1 - 4e^{3\ve\be}$ & $e^{7\ve\be}$ &
  & $1 - 6e^{2\ve\be}$ & $e^{5\ve\be}$ &
  & $1 - 6e^{3\ve\be/2}$ & $4 e^{4\ve\be}$
\\
\end{tabular}
\end{ruledtabular}
\caption{The leading-order terms in the single-phase statistical averages $p_M^\circ, p_M^\bullet$ evaluated at the transition point $\mu = \mu_t$ (adapted from Ref.~\onlinecite{MT13}).} \label{tab: p_M}
\end{table*}

The single-phase correlation factors $P_\circ$ and $P_\bullet$ are given by Eq.~\eqref{eq: P} with the averages $p_M$ replaced by $p_M^\circ$ and $p_M^\bullet$, respectively. Combined with Table~\ref{tab: p_M}, explicit expressions for $P_\circ$ and $P_\bullet$ readily follow. Namely,
\begin{subequations} \label{eq: P eval}
\begin{align}
  P_\circ &\approx e^{(6\ve-2\eta)\be}
  + 2 e^{(3\ve-\eta)\be} + 1,
\\
  P_\bullet &\approx e^{(5\ve-2\eta)\be}
  + 2 e^{(6\ve-\eta)\be} + e^{7\ve\be}
\end{align}
for the triangular lattice,
\begin{align}
  P_\circ &\approx e^{(6\ve-4\eta)\be}
  + 4 e^{(5\ve-3\eta)\be} + 4 e^{(4\ve- 2\eta)\be}
  \nn
\\
  &\quad + 2 e^{(3\ve-2\eta)\be}
  + 4 e^{(2\ve-\eta)\be} + 1,
\\
  P_\bullet &\approx e^{(3\ve-4\eta)\be}
  + 4 e^{(4\ve-3\eta)\be} + 4 e^{(5\ve-2\eta)\be}
  \nn
\\
  &\quad + 2 e^{(4\ve-2\eta)\be}
  + 4 e^{(5\ve-\eta)\be} + e^{5\ve\be}
\end{align}
for the square lattice, and
\begin{align}
  P_\circ &\approx 4 e^{(6\ve-4\eta)\be}
  + 8 e^{(9\ve/2-3\eta)\be}
  \nn
\\
  &\quad + 8 e^{(3\ve-2\eta)\be}
  + 4 e^{(3\ve/2-\eta)\be} + 1,
\\
  P_\bullet &\approx e^{(2\ve-4\eta)\be}
  + 4 e^{(5\ve/2-3\eta)\be} + 8 e^{(3\ve-2\eta)\be}
  \nn
\\
  &\quad + 8 e^{(7\ve/2-\eta)\be} + 4 e^{4\ve\be}
\end{align}
\end{subequations}
for the hexagonal lattice.

\begin{figure}
  \centering
  \includegraphics{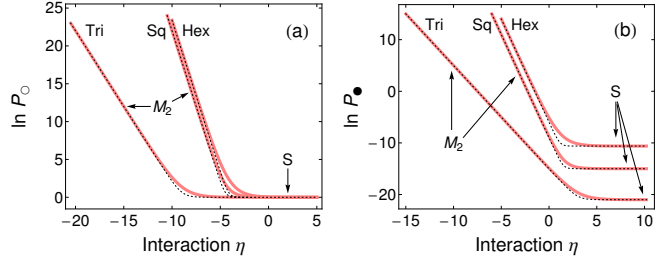}
  \caption{The dependence of the correlation factor (a) in the vacant phase and (b) in the occupied phase on the activated-state interaction $\eta$ for the three lattices (denoted as Tri, Sq, and Hex) and a fixed interaction $\ve = -3/\be$. The ranges in which the terms corresponding to the sets $M = M_2$ and $M = S$ are dominant are indicated, and the sum of these two terms for each factor and each lattice is shown by a dotted line.}
  \label{fig: P}
\end{figure}

The $\eta$ dependence of the correlation factors for a fixed $\ve$ as given by Eq.~\eqref{eq: P eval} is shown in Fig.~\ref{fig: P}. Obviously, there is a single dominant term in $P_\circ$ and $P_\bullet$ for $\eta$ either well below or well above a certain value, $\eta_\circ$ and $\eta_\bullet$, respectively. Below these vales, the dominant term corresponds to the set $M = M_2$, while above them it corresponds to the set $M = S$. On the other hand, near the values $\eta_\circ$ and $\eta_\bullet$ three or more (or even all) terms become essential. Note that $\eta_\circ = 3\ve$ and $\eta_\bullet = -\ve$ for a triangular lattice, while to $\eta_\circ = 3\ve/2$ and $\eta_\bullet = -\ve/2$ for square and hexagonal lattices. Finally, comparing the factors $P_\circ$ and $P_\bullet$ at a given $\eta$, we observe that $P_\circ$ prevails at high $\eta$, whereas $P_\bullet$ prevails at low $\eta$.

% ===============================================================
\subsection{Influence of interactions on diffusion coefficients}

Equations~\eqref{eq: D theor} -- \eqref{eq: P eval} provide explicit dependences of the diffusion coefficients $D_c$ and $D_J$ on the interactions $\ve$ and $\eta$. We may simplify these dependences and make thus their further analysis easier, if we use that for any $\eta$ only one term in $P_\circ$ and $P_\bullet$ is dominant in the coefficients.

To see this, we rewrite the two expressions in the parentheses in Eq.~\eqref{eq: D theor} as $\cP / (\te - \te_\circ)(\te_\bullet - \te)$ and $\cP / \te$, respectively, where $\cP = (P_\bullet - P_\circ) (\te - \te^*) + (P_\bullet + P_\circ) \De\te / 2$ with $\te^* = (\te_\circ + \te_\bullet)/2$. From Eq.~\eqref{eq: P} and Table~\ref{tab: p_M} we get
\begin{align} \label{eq: Pvac + Pocc}
  P_\bullet \pm P_\circ &\approx \sum_{M = M_2,S}
  e^{- (n_S - n_M) \eta\be} (p_M^\bullet \pm p_M^\circ)
  \nn
\\
  &\approx e^{- w\eta\be} p_{M_2}^\bullet \pm p_S^\circ,
\end{align}
where $w = n_S - n_{M_2}$ is equal to the number of adparticles interacting with an activated particle ($2$ for the triangular lattice and $4$ for the square and hexagonal lattices). Hence, only the terms corresponding to $M=M_2$ and $M=S$ are important in $P_\bullet \pm P_\circ$, while the other terms are exponentially suppressed (their relative contributions are of order $\exp(\ve\be/2) \ll 1$ or less). Substituting Eq.~\eqref{eq: Pvac + Pocc} into $\cP$, we readily get that the parentheses in Eq.~\eqref{eq: D theor} can be approximated by the same expressions with $P_\circ$ and $P_\bullet$ replaced by $p_S^\circ$ and $\exp(- w\eta\be) p_{M_2}^\bullet$, respectively. We thus conclude that the term $p_S^\circ$ from $P_\circ$ and the term $\exp(- w\eta\be) p_{M_2}^\bullet$ from $P_\bullet$ prevail in the diffusion coefficients at low temperatures.

Combining this with $p_S^\circ \approx 1$, $p_{M_2}^\bullet \approx \exp[(q-1)\ve\be]$, $\te_\circ \approx 0$, and $\te_\bullet \approx 1$, we get
\begin{equation} \label{eq: D theor appr}
\begin{aligned}
  D_c &\approx \frac{ D_0 e^{q\ve\be/2} }N
  \Bigl( \frac 1\te
  + \frac{ e^{ [ (q-1)\ve - w\eta] \be } }{1 - \te} \Bigr),
\\
  D_J &\approx D_0 e^{q\ve\be/2}
  \Bigl( e^{ [ (q-1)\ve - w\eta] \be }
  + \frac{1-\te}\te \Bigr),
\end{aligned}
\end{equation}
with a possible exception of coverages close to $0$ and $1$. Only two lattice parameters appear in these two formulas: $q$ and $w$, i.e., the number of adparticles interacting with a particle in the adsorbed and activated state, respectively. Note that the term $\exp(- w\eta\be) p_{M_2}^\bullet$ prevails in $D_c$ and $D_J$ when $\eta < a^* \ve$ with $a^* = (q-1)/w > 0$, i.e., when the activated-state interaction is sufficiently attractive (more than $5/2$, $3/4$, and $1/2$ of $\ve$ for the three lattices). On the other hand, if $\eta$ is repulsive or less attractive than $a^* \ve$, then the term $p_S^\circ$ prevails.

Equation~\eqref{eq: D theor appr} will allow us to easily analyze the influence of the interactions $\ve$ and $\eta$ on the diffusion coefficients. We shall discuss three specific cases.

\begin{figure}
  \centering
  \includegraphics{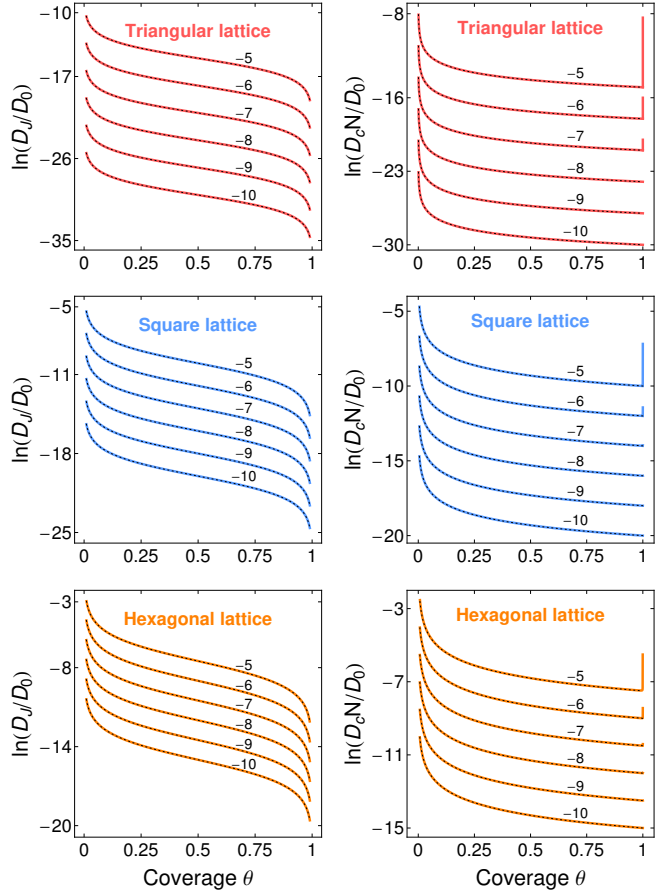}
  \caption{The coverage dependences of the diffusion coefficients for the three lattices when the activated-state interaction is neglected ($\eta = 0$). The solid lines correspond to Eq.~\eqref{eq: D theor} and the dashed lines to the approximation from Eq.~\eqref{eq: D1}. The curves are labeled by the value of $\ve\be$.}
  \label{fig: eta 0}
\end{figure}

\textbf{Case 1:} The interaction in the activated state is neglected ($\eta = 0$) and only the influence of $\ve$ occurs. Then Eq.~\eqref{eq: D theor appr} reduces to
\begin{equation} \label{eq: D1}
  D_c \approx \frac{ D_0 e^{q\ve\be/2} }N \,
  \frac 1\te \,,
  \quad
  D_J \approx D_0 e^{q\ve\be/2} \,
  \frac{1-\te}\te \,.
\end{equation}
Thus, the diffusion coefficients decrease as $\ve$ is more and more attractive (as is expected). The decrease is exponential and occurs primarily due to the chemical potential term $\exp(\mu_t\be) = \exp(q\ve\be/2)$. This behavior of the diffusion coefficients is illustrated in Fig.~\ref{fig: eta 0}. Note that the approximation from Eq.~\eqref{eq: D1} is in very good agreement with Eq.~\eqref{eq: D theor}, failing only for $D_c$ near $\te = 1$.

\emph{Remark.} When the interaction $\eta$ is neglected, only the single-site and \emph{nn}-site statistical averages occur in the correlation factor $P$. \cite{Zh85,Zh91,Zh95,Ta03,Ta07a,Ta12} Indeed, for $\eta = 0$ Eq.~\eqref{eq: P expl} reduces to $P = 1 - 2\te + \ev{\nu_1 \nu_2}$ with $\te = \ev{\nu_1}$. Then the free energy of model~\eqref{eq: H} is sufficient to evaluate $P$. However, as soon as $\eta$ is taken into account, multi-site averages $p_M$ inevitably arise, and the model's free energy cannot yield $P$. Instead, one should use the free energy of a more complex version of model~\eqref{eq: H} in which the needed multi-site interactions are added. \cite{MT13}

\begin{figure}
  \centering
  \includegraphics{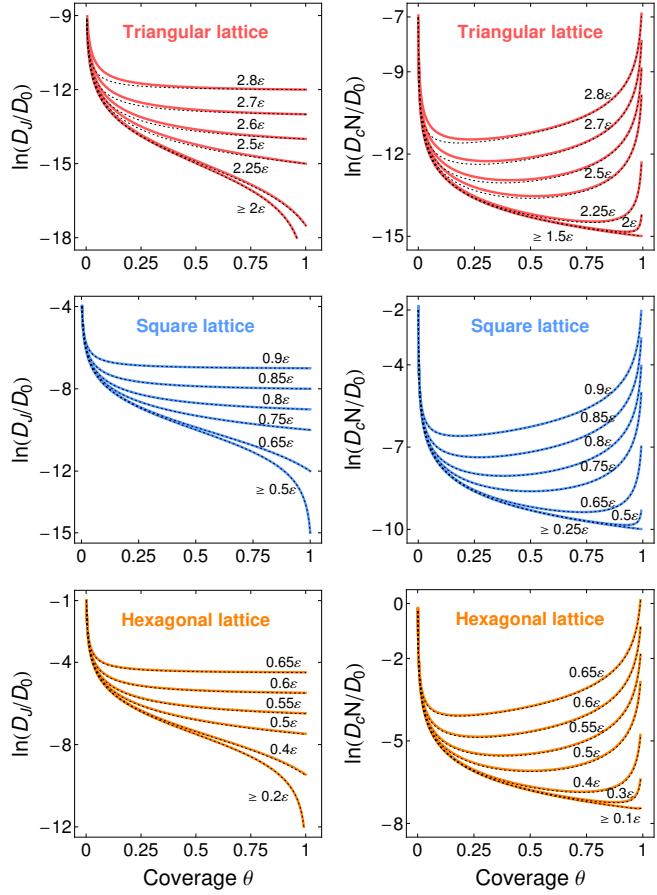}
  \caption{The coverage dependences of the diffusion coefficients for the three lattices when the adsorbed-state interaction is fixed ($\ve = -5/\be$) and the activated-state interaction $\eta$ varies. The solid lines correspond to Eq.~\eqref{eq: D theor} and the dashed lines to the approximation from Eq.~\eqref{eq: D theor appr}. The curves are labeled by the value of $\eta$. Above certain values of $\eta$ the curves are practically identical.}
  \label{fig: ep fixed}
\end{figure}

\textbf{Case 2:} The interaction $\ve$ is fixed and the value of $\eta$ varies. Then it would be anticipated that the diffusion coefficients decrease with $\eta$ repulsive and grow with $\eta$ attractive. Nevertheless, this is true only when the term $\exp(- w\eta\be) p_{M_2}^\bullet$ is dominant, i.e., when $\eta$ sufficiently attractive, $\eta < a^* \ve$. Then the diffusion coefficients grow with $\eta$ exponentially fast as $\exp(-w\eta\be)$. However, for $\eta$ repulsive or slightly attractive, $\eta > a^* \ve$, the term $p_S^\circ \approx 1$ prevails, and the diffusion coefficients are almost $\eta$ independent, $D_c \approx D_0 \exp(\mu_t\be) / N\te$ and $D_J \approx D_0 \exp(\mu_t\be) (1-\te) / \te$. Thus, no influence on surface diffusion occurs due to $\eta$, contrary to expectations, as is shown in Fig.~\ref{fig: ep fixed}. Again, the approximation of Eq.~\eqref{eq: D theor appr} is in very good agreement with Eq.~\eqref{eq: D theor}, except near $\te = 0$.

\begin{figure}
  \centering
  \includegraphics{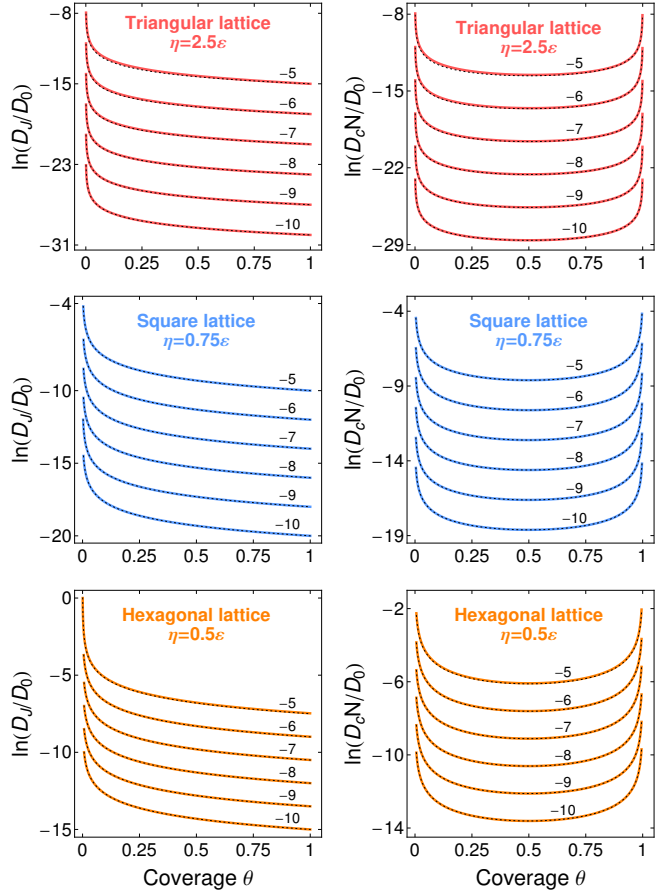}
  \caption{The coverage dependences of the diffusion coefficients for the three lattices when $\eta = a \ve$ with $a = a^* < a_0$. The solid lines correspond to Eq.~\eqref{eq: D theor} and the dashed lines to the approximation from Eq.~\eqref{eq: D3}. The curves are labeled by the value of $\ve\be$.}
  \label{fig: below a0}
\end{figure}

\textbf{Case 3.} Finally, assume that the two interactions are mutually related. For simplicity, let them be proportional to each other, $\eta = a \ve$, where $a$ is positive or negative or zero. Then Eq.~\eqref{eq: D theor appr} yields
\begin{equation} \label{eq: D3}
\begin{aligned}
  D_c &\approx \frac{ D_0 }N
  \Bigl( \frac{ e^{q\ve\be/2} }\te
  + \frac{ e^{ \la \ve \be / 2 } }{1 - \te} \Bigr),
\\
  D_J &\approx D_0
  \Bigl( e^{ \la \ve \be / 2 }
  + e^{q\ve\be/2} \, \frac{1-\te}\te \Bigr)
\end{aligned}
\end{equation}
with $\la = 3q - 2 - 2wa$. As long as $\la > 0$, the diffusion coefficients decrease with $\ve$. This happens when $a < a_0$ with $a_0 = (3q-2)/2w > a^* > 0$, i.e., when $\eta$ is either repulsive or slightly attractive relative to $\ve$. The decrease is again exponential: it behaves as $\exp(q\ve\be/2)$ for $a \leq a^*$ (the term $p_S^\circ$ prevails) and as $\exp(\la\ve\be/2)$ for $a^* \leq a < a_0$ (the term $\exp(- w\eta\be) p_{M_2}^\bullet$ prevails). We illustrate this behavior in Fig.~\ref{fig: below a0} for $a = a^*$ when both terms are important.
\begin{figure}
  \centering
  \includegraphics{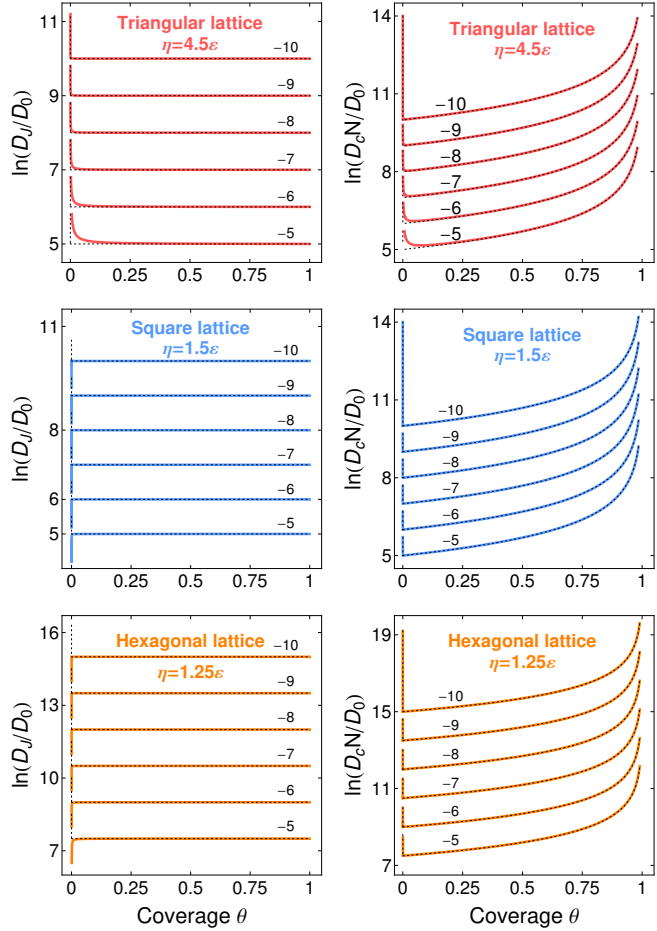}
  \caption{The coverage dependences of the diffusion coefficients for the three lattices when $\eta = a \ve$ with $a > a_0$. The solid lines correspond to Eq.~\eqref{eq: D theor} and the dashed lines to the approximation from Eq.~\eqref{eq: D3}. The curves are labeled by the value of $\ve\be$. For the square and hexagonal lattices the approximation from Eq.~\eqref{eq: D3} incorrectly increases with $\te$ near $\te = 0$.}
  \label{fig: above a0}
\end{figure}

However, if $a > a_0$, then $\la < 0$, and $\exp(- w\eta\be) p_{M_2}^\bullet$ always prevails over the term $p_S^\circ$. In this case the diffusion coefficients grow exponentially fast with $\ve$ as $\exp(\la\ve\be/2)$. Therefore, an attractive $\ve$ actually boosts diffusion as soon as $\eta$ is sufficiently attractive with respect to $\ve$---more than its $a_0$ multiple (see Fig.~\ref{fig: above a0}). For the triangular lattice the threshold value $a_0$ is as large as $4$, but for the square and hexagonal lattices it is comparable to $1$, being equal to $5/4$ and $7/8$, respectively.

% ===============================================================
\subsection{Comparison with other results}

\begin{figure}
  \centering
  \includegraphics{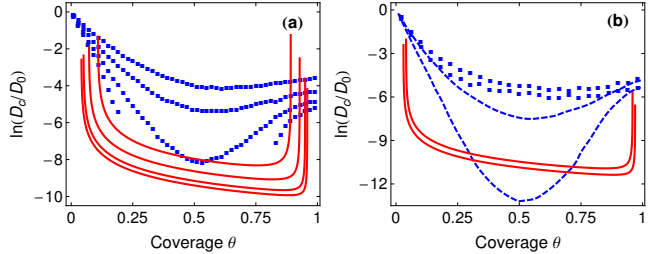}
  \caption{The chemical diffusion coefficient with the interaction in the activated state neglected ($\eta = 0$) for (a) a triangular lattice and temperatures $T = - K \ve / k_B$ with $K = 1.35, 1.14, 1.0, 0.94$ from top to bottom (adapted from  Ref.~\onlinecite{Ta03}); (b) a square lattice and temperatures $T = - K \ve / k_B$ with $K = 0.63$ (upper) and $K = 0.57$ (lower) (adapted from Ref.~\onlinecite{Ta00}). Squares represent Monte Carlo simulations, dashed lines a renormalization group method, and solid lines Eq.~\eqref{eq: D theor}.}
  \label{fig: compar}
\end{figure}

As mentioned in Section~\ref{sec: MODEL}, model~\eqref{eq: H} was previously used to study surface diffusion on a triangular and square lattices, using Monte Carlo simulations and real-space renormalization group methods. \cite{Ta00,Ta01,Ta03} However, temperatures above the critical point were considered there, so a meaningful comparison with our low-temperatures results cannot be carried out. To illustrate this, in Fig.~\ref{fig: compar} we show the previously obtained results as well as our results for the same temperatures in the case when $\eta$ is neglected, boldly applying Eq.~\eqref{eq: D theor} in the over-critical region. Clearly, the agreement is only a qualitative and quickly worsens as the temperature grows.

% ####################################################### SECTION
\section{Conclusions}

We investigated the interplay between the opposing effects of the interactions in the adsorbed and activated states on surface diffusion at subcritical temperatures. To this end, the coverage dependences of the chemical and jump diffusion coefficients $D_c$ and $D_J$ were considered and their sensitivity to the interactions was studied. We stressed that at low temperatures a phase transition is necessary whenever a coverage dependence of any quantity is to be explored; far from transitions a single phase is stable in the system and the coverage is almost constant.

We considered a simple model in a finite portion of a triangular, square, and hexagonal lattice with periodic boundary conditions. An adsorbed particle was assumed to interact only with its $q$ nearest neighbors via an attractive energy $\ve$, while an activated particle interacted with its $w$ nearest adparticles via a different (attractive or repulsive) energy $\eta$. The model is known to exhibit a first-order phase transition at low temperatures between the fully vacant and fully occupied phase. Uncorrelated \emph{nn} jumps were assumed. Thus, the activated-state interaction $\eta$ was associated with segments $S$ of $w+2$ lattice sites, two of which represented a \emph{nn} pair of vacant sites between which a jump was performed.

For this model we were able to explicitly evaluate the dependence of $D_c$ and $D_J$ on the two interactions. It turned out that only two contributions were dominant, one for each phase: in the vacant phase it corresponded to the segment $S$ in which all $w+2$ sites were vacant; in the occupied phase it corresponded to $S$ with the maximal number $w$ of occupied sites. Three specific cases were discussed.

First, if $\eta$ was neglected, then, as expected, $\ve$ decelerated diffusion. We showed that the rate of deceleration was exponential, as $\exp(q\ve/2k_B T)$.

Second, if $\ve$ was fixed and $\eta$ varied, then an attractive $\eta$ accelerated diffusion only when it was more attractive than $(q-1)/w \times \ve$; again, the rate of acceleration was exponential, this time as $\exp(-w\eta/k_B T)$. On the other hand, a less attractive or repulsive $\eta$ had practically no effect on diffusion.

The most intriguing case occurred when the two interactions were considered to be proportional, $\eta = a \ve$. Then surface diffusion was exponentially accelerated (decelerated) whenever $a > a_0$ ($a < a_0$), where the threshold value $a_0 = (3q-2)/2w$. Thus, an $\eta$ comparable to $\ve$ is enough to boost diffusion on the square and hexagonal lattices ($a_0 = 5/4$ and $a_0 = 7/8$), while on the triangular lattice $\eta$ must be more than four times as attractive as $\ve$.

The key ingredient underlying our investigation was Eq.~\eqref{eq: D theor}. It allows one to study the case of any relation, $\eta = f (\ve)$, between the two interactions; the linear relation $\eta = a \ve$ was considered due to its simplicity. Moreover, since the equation is of rather general nature, \cite{MT12a} the approach presented here for model~\eqref{eq: H} can be applied also to other models on homogeneous or even heterogeneous lattices. This only requires to calculate the single-phase correlation factors and coverages for the studied model and at a given phase transition, which is simple to carry out heuristically, as was shown in Section~\ref{sec: RES}. A rigorous evaluation can be done via low-temperature cluster expansions. \cite{MT13}

% ####################################################### SECTION

\section*{Acknowledgments}

This research was supported by the Czech Science Foundation, Project No.~P105/12/G059.

% ####################################################### SECTION

%\bibliographystyle{apsrev}
%\bibliography{SD-REF}

\end{document}